\documentclass[a4paper,fleqn]{cas-dc}
\makeatletter
\def\@thefoot{\hfil\thepage\hfil}
\def\@oddfoot{\@thefoot}
\makeatother

\usepackage[authoryear,longnamesfirst]{natbib}
\usepackage{expl3}
\usepackage{graphicx}
\usepackage{subfig}

\def\tsc#1{\csdef{#1}{\textsc{\lowercase{#1}}\xspace}}
\tsc{WGM}
\tsc{QE}

\begin{document}
\let\WriteBookmarks\relax
\def\floatpagepagefraction{1}
\def\textpagefraction{.001}

\shorttitle{}    

\shortauthors{}  

\title [mode = title]{Tetrahedron-Net for Medical Image Registration}   



%

\author[1]{Jinhai Xiang}
\credit{Writing–review \& editing, Supervision, Resources, Conceptualization, Funding acquisition}
\author[1]{Shuai Guo}
\credit{Writing–original draft, Methodology, Visualization, Validation}
\author[1]{Qianru Han}
\credit{Methodology, Investigation}
\author[1]{Dantong Shi}
\credit{Writing–original draft, Investigation}
\author[1]{Xinwei He}
\cormark[1]
\credit{Writing–review \& editing, Project administration, Formal analysis}
\author[2]{Xiang Bai}
\credit{Writing–review \& editing}






\affiliation[1]{organization={Huazhong Agricultural University},
            city={Wuhan},
            postcode={430070}, 
            country={China}}




\affiliation[2]{organization={Huazhong University of Science and Technology},
            city={Wuhan},
            postcode={430074}, 
            country={China}}

\cortext[1]{Corresponding author}



\begin{abstract}
Medical image registration plays a vital role in medical image processing.
Extracting expressive representations for medical images is crucial for improving the registration quality. 
One common practice for this end is constructing a convolutional backbone to enable interactions with skip connections among feature extraction layers.
The de facto structure, U-Net-like networks, has attempted to design skip connections such as nested or full-scale ones to connect one single encoder and one single decoder to improve its representation capacity.
Despite being effective, it still does not fully explore interactions with a single encoder and decoder architectures.
In this paper, we embrace this observation and introduce a simple yet effective alternative strategy to enhance the representations for registrations by appending one additional decoder.
The new decoder is designed to interact with both the original encoder and decoder. In this way, it not only reuses feature presentation from corresponding layers in the encoder but also interacts with the original decoder to corporately give more accurate registration results.
The new architecture is concise yet generalized, with only one encoder and two decoders forming a ``Tetrahedron'' structure, thereby dubbed Tetrahedron-Net. 
Three instantiations of Tetrahedron-Net are further constructed regarding the different structures of the appended decoder. 
Our extensive experiments prove that superior performance can be obtained on several representative benchmarks of medical image registration. Finally, such a ``Tetrahedron'' design can also be easily integrated into popular U-Net-like architectures including VoxelMorph, ViT-V-Net, and TransMorph, leading to consistent performance gains.
\end{abstract}



\maketitle

\section{Introduction}
\begin{figure}[h]
    \begin{center}
    \centerline{\includegraphics[width=1\linewidth]{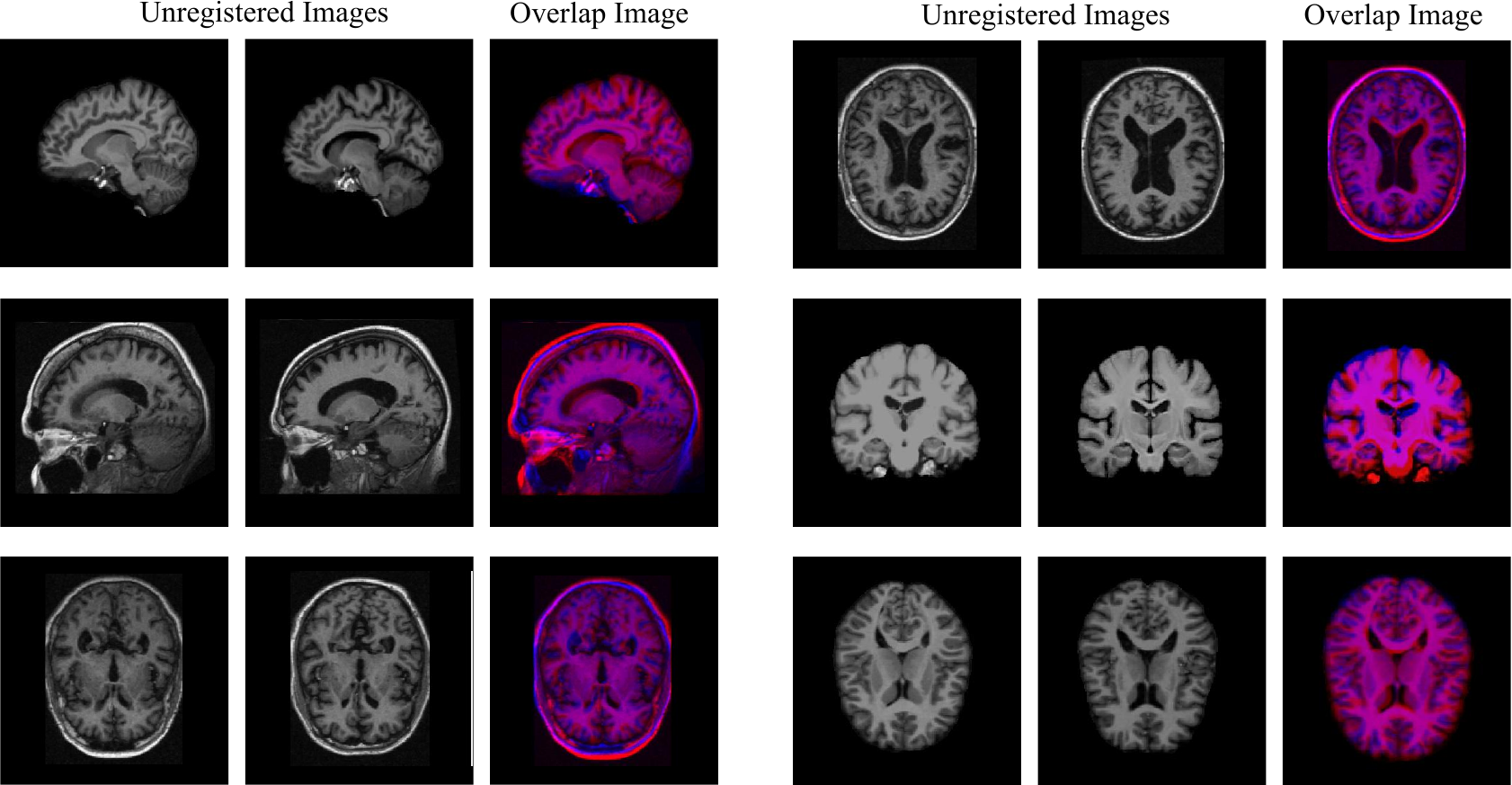}}
    \caption{Visualization of unregistered image pairs and the overlapping images. Overlapping the two images with different colors suggests that there are significant mismatches in the image pairs.}
    \label{x_to_y}
    \end{center}
\end{figure}

Medical image registration (MIR) aims to accurately align one source medical image relative to a fixed target one depicting the same underlying anatomical structures. 
It is a crucial processing step for a variety of clinical applications such as image-guided surgical treatment~\cite{imageguided}, disease diagnosis~\cite{review2021}, and disease progress monitoring~\cite{review2018}. However, MIR is an extremely challenging task because the two medical images are generally taken from different viewpoints or temporal phases, and the same anatomical structures typically exhibit distinct shapes and appearances (see Fig.~\ref{x_to_y}).

A large body of works~\cite{review2023,2022review} has been presented to address this task. 
Conventional medical image registration methods, \emph{e.g.}, elastic~\cite{elastic}, fluid~\cite{fluid}, or B-spline models~\cite{b-spline}, formulate this task as an optimization problem that learns to maximize the appearance similarity while assuring regular transformation between the moving source and fixed target images. 
They generally suffer from high computational complexity and slow convergence because of the necessity of applying the optimization process for each pair of images. 
Another promising line is deep neural network-based approaches.
These methods generally formulate MIR as a deep regression problem on registration parameters: displacement fields, velocity fields, and momentum fields. During training, a convolutional neural network is trained end-to-end to output the registration parameters for a source image with respect to the target one. Once the \emph{data-driven} training phase is completed, the registration parameters can be obtained directly with only one forward pass, thereby greatly improving the speed and reliability. 

In recent years, MIR has been greatly advanced by the progress of backbone architectures.
Popular convolutional architectures such as FCN~\cite{fcn}, GoogleNet~\cite{googlenet}, and ResNet~\cite{resnet} have been widely adopted.
Among them, U-Net-like ones are predominant in many state-of-the-art medical image registration frameworks~\cite{review2023}, which demonstrates superior multi-scale learning capacity with skip connections in a symmetric encoder-decoder structure.
Inspired by their promising performance, some works have also attempted to redesign skip connections. For instance, UNet++~\cite{zhou2019unet++} designs nested and dense skip connections to derive a built-in ensemble of UNets of varying depths. 
UCTransNet~\cite{UCTransNet} replaces the skip connections with a Channel Transformer module. On the other hand, the advances in general backbone networks such as DenseNet~\cite{huang2017densely}, ViT~\cite{dosovitskiy2020image} have also been readily incorporated into U-Net to produce strong U-Net. For instance, Transformer-UNet~\cite{wang2022transformer} integrates ViT into UNet and advances the state-of-the-art greatly. However, the aforementioned works, which adhere to the U-Net family, may be constrained by the inherent limitations of the U-Net architecture itself. The traditional U-Net design, characterized by a single encoder-decoder pair, may not fully capture the complexity and diversity of features required for advanced medical image registration tasks. Furthermore, the practice of excessively adding dense and sophisticated skip connections can introduce unnecessary complexity, potentially leading to redundancy and diminishing returns in terms of performance gains.

In this work, we present a simple yet efficient architecture named Tetrahedron-Net for medical image registrations. Compared with U-Net-like architectures, our framework has one additional decoder branch that enhances the registration parameter decoding capacity. In other words, our decoder is indeed a two-level decoder: one level aims to effectively collect rich representations from the encoder and output coarsely decoding results, and the other attempts to refine the coarse results by connecting the encoder and the first level decoder with skip connections. By incorporating the two-level decoding process, the registration parameters can be easily and accurately regressed in a coarse-to-fine manner than traditional UNets making one straightforward prediction. 

Our framework is a concise yet effective extension to classic UNet. 
Compared with other U-Net-like variants which adopt two parallel yet separate branches~\cite{huang2022tdd}, the two-level decoding processing can learn the registration parameters cooperatively.
When compared UNet++~\cite{zhou2019unet++}, our design does not require additional prune methods during inference, which may cause architecture bias between training and inference, leading to performance degradation.
With such a two-level decoder, Tetrahedron-Net reaches a new state-of-the-art performance on medical image registration.

Notably, the above physiology of designing a two-level decoder can be further extended to other popular U-Net-like architectures and further boost their performance. We replace the decoder with a two-level decoder on four architectures covering VoxelMorph~\cite{Voxelmorph}, ViT-V-Net~\cite{chen2021vit}, TransMorph~\cite{transmorph}, and TransMorph-bspl~\cite{transmorph}. All display consistent improvements. 

To summarize, the main contributions of our method are as follows: 
1) We present a simple yet effective framework named \textbf{Tetrahedron-Net} for medical image registration. Its core component is a two-level decoder, with one level connecting the encoder and the second-level connecting both the encoder and first decoder, thereby corporately regressing the registration parameters accurately. Extensive experiments are conducted on three MIR benchmarks, proving its effectiveness. 
2) The proposed two-level decoding physiology is general and effective. We further apply it to four U-Net-like architectures and observe notable gains coinsistently. 

\section{Related Work} 
\subsection{Unsupervised Image Registration}
Traditional supervised learning methods rely heavily on the quality of the gold standard. 
STN (spatial transformer network) was proposed~\cite{stn}, which allows networks to implement spatial transformations on moving images based on deformation fields. 
It can be directly inserted into existing convolutional registration networks, making it possible to compute the loss of image similarity during the training process. 
This network has pushed the unsupervised registration research with the following optimization objective:
\begin{equation}
    \phi = argmin \; \mathcal{L}_{sim}(I_{m}^{warped},I_{f})
\end{equation} 
where $I_{m}$ represents moving image; $I_{f}$ represents fixed image; $\phi$ represents the deformation field; $L_{sim}$ represents the similarity between $I_{m}^{warped}$ and $I_{f}$. 

DIRNet~\cite{dirnet} was the first unsupervised registration network based on image similarity, using the similarity between $I_{m}^{warped}$ and $I_{f}$ as the loss function, making end-to-end network training possible.
For instance, VoxelMorph~\cite{Voxelmorph} used a CNN architecture similar to UNet~\cite{unet} to acquire the deformation field. VoxelMorph has achieved wide acceptance in the field of medical image registration, with considerable improvements in registration speed and accuracy. Moreover, an explicit penalty loss computing negative Jacobian determinants is used to extend VoxelMorph, named as FAIM~\cite{faim}. 
However, these approaches might not accurately estimate large displacements within complicated deformation fields. To tackle this challenge, recent developments focus on employing a series of stacked networks. Zhao et al. crafted a recursive cascading network where multiple VoxelMorph networks are layered recursively to progressively warp the images~\cite{DataAU}. Kim et al. introduced CycleMorph~\cite{CycleMorph}, which consists of two registration networks, taking inputs by switching their orders with a cycle consistency - this innovation allows the model to more effectively grasp transformation relationships across various levels, yet it comes with high complexity and computational. 
In order to address the relatively limited ability of convolutional networks to understand spatial relationships over long distances in images, ViT-V-Net~\cite{chen2021vit,vitreview} integrates a Vision Transformer block after encoder. TransMorph~\cite{transmorph,transformersreview, Transforming} is a perfect blend of Transformer and ConvNet, taking full advantage of the strengths of both. the structure of TransMorph is the same as the classic UNet structure but with the innovative fusion and enhancement of the Swin Transformer~\cite{swintransformer} at each layer of the network in the encoder.

\subsection{Registration network network based on u-shaped network}
Most of the current registration models are still designed based on UNet structures. In addition to the u-shaped networks already mentioned in the previous section, for example, ICNet ~\cite{icnet} developed an inverse consistency constraint that deforms an input pair of images symmetrically towards each other until the two deformed images reach a matching state; U-ReSNet ~\cite{u-ReSNet} constructed an encoder-decoder-like network for brain MRI image registration; and VTN (Volume tweening network) ~\cite{vtn} was proposed with an additional internal affine pre-registration module to optimize the performance of the network in terms of deformable registration, based on this ~\cite{2023reinforcement}, they further constructed an end-to-end recursive cascaded network RCVTN (recursive cascaded VTN)~\cite{rcvtn}, which learns complex spatial mapping relationships accurately and progressively through cascading operations of multiple sub-networks. These studies have predominantly concentrated on enhancing the encoder of UNet, but overlooked the significance of the UNet decoder.

With the continuous promotion of the U-shape network, a large number of more novel model structures have emerged. Compared with the traditional UNet model, UNet ++ ~\cite{zhou2019unet++} successfully enhances the feature expression efficiency of the network by using the multi-resolution feature fusion and skip connection structure as cleverly as possible, thus optimizing the output of the model more efficiently. Attention-UNet ~\cite{attentionunet} introduces an attention mechanism, which enables the network to automatically focus on important features during the learning process, thus improving the recognition performance of the target region. UNet 3+~\cite{unet3+} is another improvement of UNet, by adding additional paths and modules to UNet, it effectively expands the perceptual range of the network and improves the feature expression capability, thus improving the accuracy of the network. DenseUNet~\cite{denseunet,2022unsupervised} combines the structure of dense connectivity and UNet, which greatly facilitates the flow of information and feature transfer by tightly connecting the output of each layer to the output of all previous layers, thus improving the learning ability of the network. These different UNet morphing networks have supported and assisted us in our improvements.

\section{Method}
\subsection{Problem Formulation and Motivation}
Following previous popular works such as Voxelmorph ~\cite{Voxelmorph} and LKU-Net ~\cite{Jia2022UNetVT}, we formulate medical image registration as the deformation field prediction problem. 
Given a training set of $N$ medical image pairs $\mathcal{T} = \{(f^i, m^i)\}_{i=1}^N$, where $f^i$ and $m^i$ denotes fixed and corresponding moving images respectively, our main goal is to learn a network $F(\cdot)$ to predict dense deformation field $\phi$ which maps the coordinates from $f$ to $m$ with the following objectives:
\begin{equation}
    \text{min} \; \mathcal{L}(f^i, m^{i} \circ \phi^i),
\end{equation} 
where $\phi^i = F(f^i, m^i)$, and $m^{i} \circ \phi^i$ denotes warping $m^{i}$ with $\phi^i$. 
During testing, the smooth deformation field can be obtained with a single forward pass by feeding each test medical image pair $(f^{te}, m^{te})$ into the network: $\phi^{*} = F(f^{te}, m^{te})$, and then warp the moving image with the predicted field. 

Note that the deformation field prediction network $F(\cdot)$ can be instantiated with any off-the-shelf convolutional neural networks. 
For dense prediction tasks, multi-scale contextual information is crucial. Previous works have demonstrated that UNet is very effective at predicting dense deformation fields. With a symmetric encoder-decoder structure connected by skip connections, it has a strong capacity to extract features at different granularities.
However, most works rely on one encoder to encode the image pairs and then decode it once. It poses a great challenge in handling cases when moving images have large relative displacement with respect to the fixed one. Zhao et al.~\cite{zhao2019recursive} further introduce recursively apply UNet to address this issue. However, it increases the computation burden and is less efficient. To address this issue, we present an alternative way by extending UNet with a two-level decoder while keeping one shared encoder.
The two-level decoder is composed of two coupled decoders that work cooperatively to predict the deformation field. In this way, our framework makes the best of the first-level decoder results as a prior and facilitates the more refined prediction by the second-level decoder, thereby better handling the challenging large deformation scenarios. 

\begin{figure*}[htbp]
\centerline{\includegraphics[width=1\linewidth]{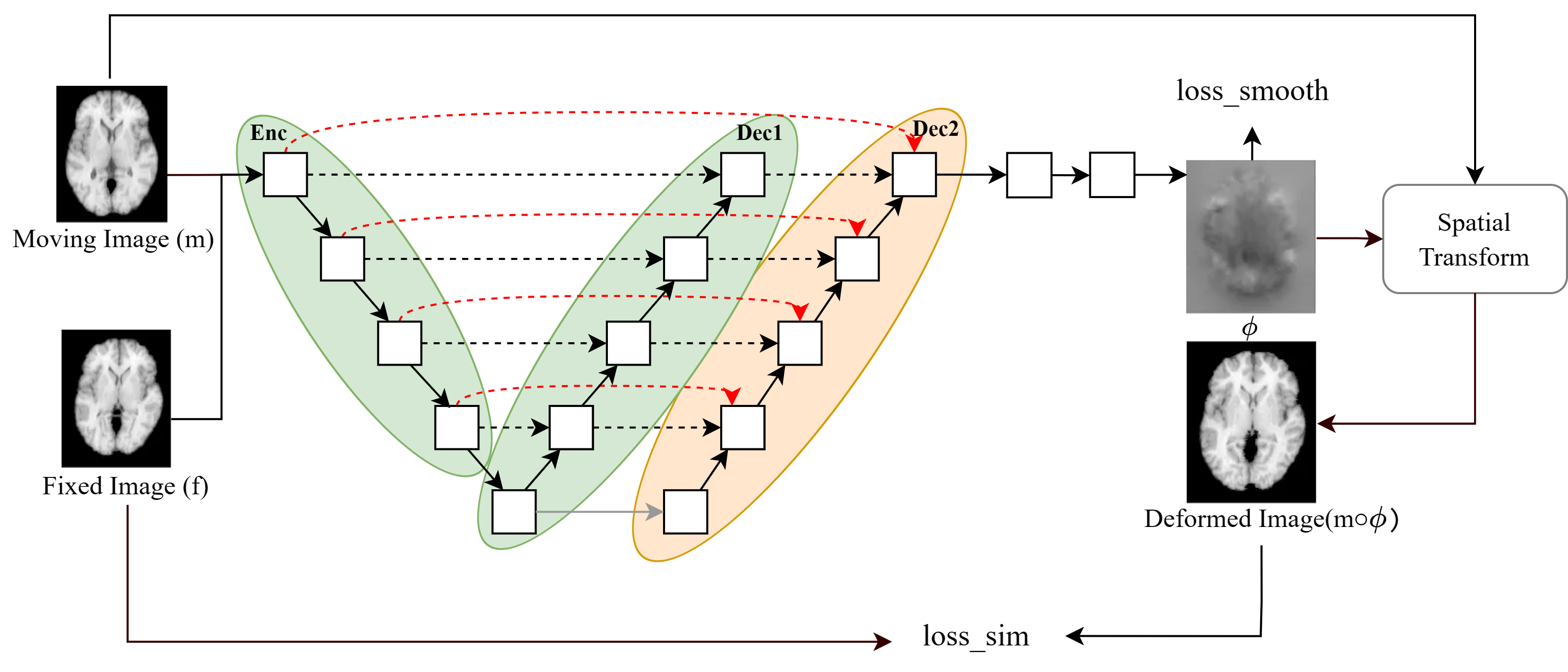}}
\caption{The architecture of the proposed Tetrahedron-Net registration network. The registration network used in the figure is the U-UNet network. 
Firstly, the fixed image $f$ and the moving image $m$ are concated in the channel dimension as the input. After the encoder extracts the features and then two decoders generate the deformation field $\phi$. Then the spatial transformation network (STN) uses the generated deformation field $\phi$ to deform the moving image $m$ to obtain the Deformed image ($m\circ \phi$), the loss of smoothness (loss\_smooth) is calculated for the generated deformation field, and the loss of similarity (loss\_sim) is calculated for the generated deformed and fixed images. The structure of the encoder(Enc) and Decoder1(Dec1) as same as UNet, and the Decoder2(Dec2) with the same structure as Dec1. The circles in the figure represent the concat operation and the squares represent two consecutive $3 \times 3 \times 3$ Convolution and ReLU layers.}
\label{u-registration}
\end{figure*}
\subsection{Tetrahedron-Net}
\subsubsection{Overview} 
Fig.~\ref{u-registration} gives an overview of Tetrahedron-Net. As shown, it consists of three key components, \emph{i.e.}, an encoder (Enc), 1st-level decoder (Dec1), and 2nd-level decoder (Dec2).
The inputs of our framework are fixed and moving image pairs, denoted by $f \in \mathbb{R}^{C \times H \times W \times D}$ and $m \in \mathbb{R}^{C \times H \times W \times D}$, respectively, where $C$, $H$, $W$ and $D$ represent channel, height, width and depth of the image. 
We first concatenate them along the channel dimension $g \in \mathbb{R}^{2C \times H \times W \times D}$ and feed them into the encoder. The encoder is composed of several convolutional layers and progressively downsamples the input to a set of hierarchical feature maps. 
After extracting the set of multi-scale representations, we further feed them into the corresponding decoder layers with skip connection to predict the deformation field $\phi  \in \mathbb{R}^{H \times W \times D}$. Our decoder is a two-level decoder. The 1st-level decoder aims to combine the multi-scale representations from the encoder,  and the 2nd-level decoder further refines the features from the 1st-level decoder with the encoder representations, thereby facilitating deformation field prediction in a coarse to fine manner. 
Finally, the registration image can be easily obtained by applying spatial transformation network (STN)~\cite{stn} to deform the moving image based on the predicted deformation field ($m\circ\phi$).   
The whole framework is optimized with image similarity loss between the fixed and the moving image under smoothing constraints over the deformation field. 

\subsubsection{Encoder}
The encoder of Tetrahedron-Net is designed with the same philosophy as UNet, which comprises a series of blocks to perform feature extraction, decreasing the feature map resolution progressively. 
Each block in the encoder is simply composed of two $3 \times 3 \times 3$ Convolution and ReLU layers (denoted by CR($\cdot$)), followed by a max-pooling layer with a $2 \times 2 \times 2$ window and stride 2, reducing the resolution by a factor of 2.
Mathematically, given an input tensor $g \in \mathbb{R}^{2C \times H \times W \times D}$, the computation process of the encoder is formulated as follows: 
\begin{equation}
\label{equ_3}	
	\begin{aligned}	
		x_{enc}^{i}= 
		&\left\{\begin{aligned}	
		&g &  & i=0\\
		&\text{Maxpool}(\text{CR}(x_{enc}^{i-1}))& & i\in \left [ 1, L \right ] \\
	\end{aligned}\right.\\
	\end{aligned}
\end{equation}
where $L$ is the number of scales, and $\{x_{enc}^{i}\}_{i=1}^{L}$ represent representations of cascaded blocks for achieving the full resolution $H \times W \times D$ to low resolution $\frac{H}{2^{L}} \times \frac{W}{2^{L}} \times \frac{D}{2^{L}}$.  

In addition to the original encoder from UNet, any off-the-shelf encoder can be utilized here. 
Our framework can also be integrated with more sophisticated encoders such as ViT-V-Net and TransMorph, which can further improve the registration accuracy based on our empirical results. 

\subsubsection{Two-Level Decoder}
Our decoder is a two-level decoder that works cooperatively. The first-level decoder works exactly the same way as in the original UNet. It consists of a stack of convolutional blocks that are applied to gradually learn to upsample from $\frac{H}{2^{L}} \times \frac{W}{2^{L}} \times \frac{D}{2^{L}}$ to the full resolution $H \times W \times D$. 
Mirroring the encoder, each block in the decoder is connected with the corresponding one in the encoder part to make use of low-level details, thus further facilitating resolution recovery. Formally, the above process is formulated as follows:
\begin{equation}
\label{equ_4}	
	\begin{aligned}	
		x_{dec1}^{j}=
		&\left\{\begin{aligned}	
		&\text{Up}(x_{enc}^{L-j})&  & j=0 \\
		&\text{Up}(\text{CR}([x_{enc}^{L-j},x_{dec1}^{i-1}]))&  & j\in \left [1,L \right ]  \\
	\end{aligned}\right.\\
	\end{aligned}
\end{equation}
where $\text{Up}(\cdot)$ represents the transpose convolution for upsampling, $j$ represents block index in the decoder, and [$\cdot$] indicates concatenation operation. 

At the second level, another decoder branch makes use of representations from the encoder and the first-level decoder via skip connections to predict the deformation fields. 
\begin{equation}\label{equ_5}	
	\begin{aligned}	
		x_{dec2}^{k}=
		&\left\{\begin{aligned}	
		&\text{Up}(x_{enc}^{L-k})&  & k=0\\
		&\text{Up}(\text{CR}([x_{enc}^{L-k},x_{dec1}^{k-1},x_{dec2}^{k-1}]))&  &
k\in \left [1,L \right ]   \\
	\end{aligned}\right.\\
	\end{aligned}
\end{equation}
where $x_{dec2}^{*}$, $x_{enc}^{*}$ represents the outputs from specific encoder and decoder blocks which are indexed by subscript *, respectively. 
In this way, the model can comprehensively utilize both the low- and high-level features for fusion to obtain better registration results. 
Note that in our framework, the first-level decoder works cooperatively with the second-level since it provides prior information for the second-level decoder, enabling our framework to accurately handle complex large deformation in a coarse to fine manner.

\subsection{Various Second-Level Decoder Structures}
Tetrahedron-Net can be treated as a modified framework by appending a second-level decoder (Dec2) to the UNet. Therefore the architecture design for the second-level decoder is flexible.  
To further demonstrate the generalizability, we further design three representative structures for the second-level decoder, including the UNet++~\cite{zhou2019unet++}, UNet3+~\cite{unet3+}, and DenseUNet~\cite{denseunet}. 

\subsubsection{UNet++} 
\begin{figure}[htbp]
\setlength{\belowcaptionskip}{-0.5cm} 
\centerline{\includegraphics[width=1\linewidth]{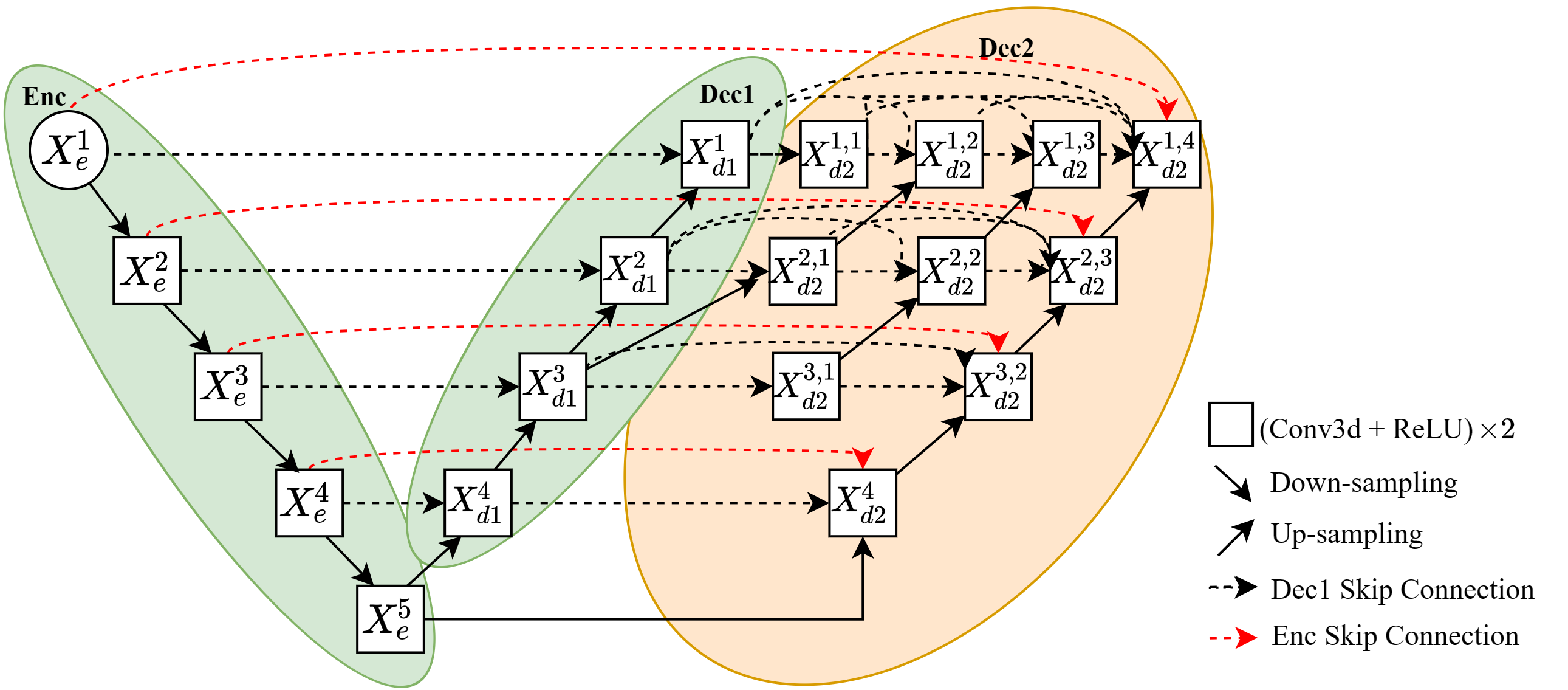}}
\caption{U-UNet++ is adopted as the second-level decoder. Each node is composed of convolution and ReLU layers.}
\label{u-uplus}
\end{figure}

As shown in Fig.~\ref{u-uplus}, the second-level decoder is designed following UNet++. In this configuration, each decoder layer is connected to all preceding decoder layers. We realize the upsampling layer with transposed convolution (ConvTranspose3d), ensuring fidelity in feature reconstruction.
Compared to U-UNet, adopting U-UNet++ as the second-level decoder can better capture the detail information and context information, and this dense connectivity mechanism enhances the transfer and reuse of the features of decoder 1 (Dec1), which can improve the expressive power of the network and the stability.

\subsubsection{UNet3+} 
\begin{figure}[htbp]
\centerline{\includegraphics[width=0.95\linewidth]{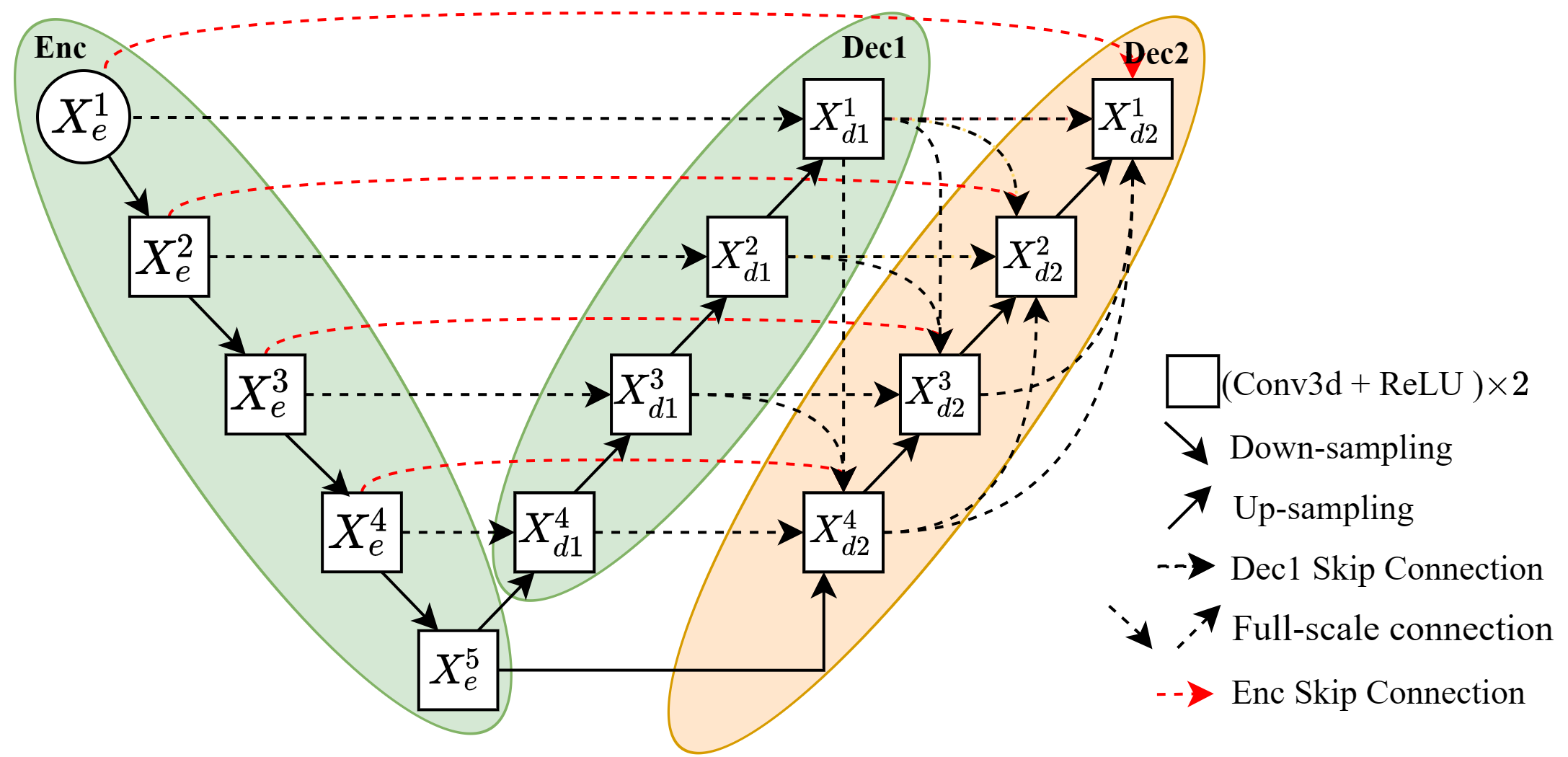}}
\caption{The second-level decoder is designed by U-UNet3+.}
\label{u-u3plus}
\end{figure}

Fig.~\ref{u-u3plus} shows the second-level decoder structure in the form of UNet3+.
As shown, each layer within Decoder 2 (Dec2) integrates feature maps from all scales of Decoder 1 (Dec1) as well as larger-scale feature maps from its own structure. Note that feature maps at the same scale are directly fused, while for deeper feature maps with small resolutions from Dec1, we upsample them to match the resolution. Otherwise, we downsample them for shallow features with max-pooling. It allows both fine-grained and coarse-grained features to be captured, resulting in clearer boundary delineation and consequently enhancing overall accuracy.

\subsubsection{U-DenseUNet} 
\begin{figure}[htbp]
\centerline{\includegraphics[width=0.95\linewidth]{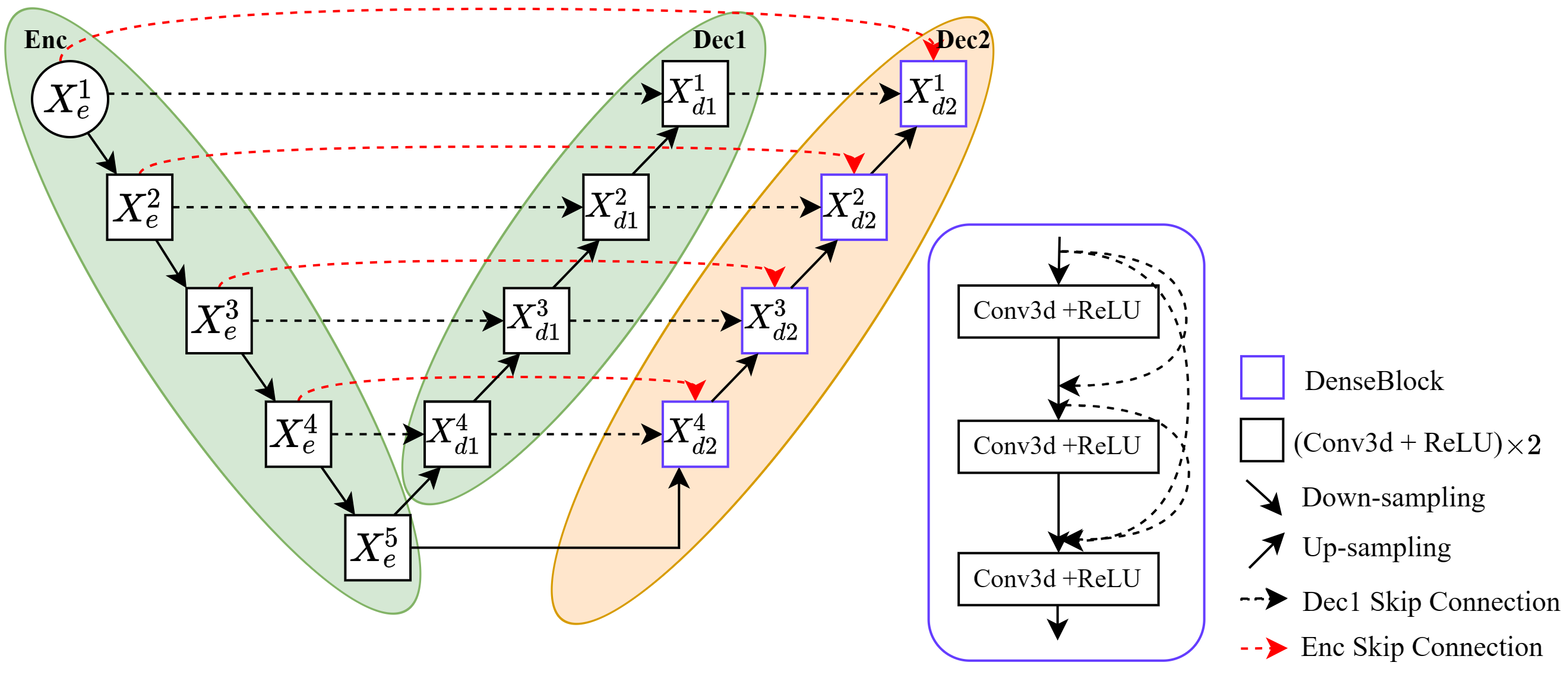}}
\caption{The second-level decoder is designed by U-DenseUNet.}
\label{u-denseu}
\end{figure}

As shown in Fig.~\ref{u-denseu}, the second-level decoder takes the form of DenseUNet. Here each decoding layer is simply a DenseBlock that has dense connections across all layers. Such a design strategy encourages feature reuse and feature propagation, yielding a framework more easier to train and inducing more accurate registration parameters.

\subsection{Skip Connections} The encoder features are fused using the skip connection mechanism, and for the different Dec2s mentioned above, we used different skip connections. For U-UNet and U-DenseUNet, we directly use concat to fuse the encoder features of the same layer with those of Dec2, 
whereas for U-UNet++ and U-UNet3+, concatenating the features from the encoder
after its full fusion of Dec1's features.

\subsection{Integration with U-Net-like Architectures}
Since the seminal work UNet, a great number of UNet variants have been presented to further improve its capacity in learning multi-scale representation for dense predictions. Their success has been nicely transferred to medical image registration.
Since the two-level decoder does not put any constraints on the overall network structure, we can easily integrate a two-level decoder into any off-the-shelf architectures and further improve their capacity. 
To show its generality, we select three representative U-Net-like backbones including  VoxelMorph~\cite{Voxelmorph}, Vit-V-Net~\cite{chen2021vit}, TransMorph~\cite{transmorph}, and then replace the decoder with the two-level decoders to derive the Tetrahedron-Net like architecture.

\subsection{Loss Function}
The training objective of our framework $\mathcal{L}_{final}$ comprises two terms: one is the similarity loss $\mathcal{L}_{sim\_img}$ defined on fixed image $f$ and the warped images that are obtained by applying deformation field $\phi$ on moving images $m$, and the other is the regularization term $\mathcal{L}_{smooth}$ to smooth the deformation fields. Mathematically, it is formulated as follows:
\begin{equation}\label{total}
\begin{split}
    \mathcal{L}_{final}(I_{f},I_{m},\phi)=\mathcal{L}_{sim\_img}(I_{f},I_{m} \circ \phi)+\lambda\mathcal{L}_{smooth}(\phi)
\end{split}
\end{equation}
where $\circ$ represents the deformation operation. 
Normalized Cross Correlation (NCC) is used to calculate image similarity, which is defined as follows:
\begin{equation}
\begin{split}
&\mathcal{L}_{sim\_img}(I_{f},I_{m},\phi )= \\
&\sum_{p\in \Omega}\frac{ (({\textstyle \sum_{p_{i}}I_{f}(p_{i})-\widehat{I}_{f}(p))([I_{m}\circ\phi ](p_{i})-[\widehat{I}_{m}\circ\phi](p_{i})))^{2} }}
               { ({\textstyle \sum_{p_{i}}(I_{f}(p_{i})-\widehat{I}_{f}(p))^{2})( {\textstyle \sum_{p_{i}}([I_{m}\circ\phi ](p_{i})-[\widehat{I}_{m}\circ\phi](p))^{2})} )}} 
\end{split}
\end{equation}
The regularization term is implemented to smooth the deformation field:
\begin{equation}
    \mathcal{L}_{smooth}(\phi) = \sum_{p\in \Omega }\left | \left | \bigtriangledown \phi (p) \right |  \right | ^2
\end{equation}

\section{Experiments}
\subsection{Experimental Setup}
\subsubsection{Datasets and Preprocessing} 
This study employs four public datasets convering LPBA 40~\cite{LPBA40}, IXI~\cite{ixi}, and OASIS~\cite{OASIS}. For all datasets, we conduct standard preprocessing procedures on structural brain MRI data using FreeSurfer~\cite{FreeSurfer}, including skull stripping, resampling, and affine transformation.
For LPBA40, the volume dimension is 160 $\times$ 192 $\times$ 160, we used 30 volumes for training, 9 volumes for testing, and 1 volume as the atlas. For IXI, the volume dimension is 160 $\times$ 192 $\times$ 224, and it was split into 403, 58, and 115 (7:1:2) volumes for training, validation, and test sets. For OASIS, the volume dimension is 160 $\times$ 192 $\times$ 224, and 394, 19, and 38 images are being used for training, validation, and testing.

\subsubsection{Implementation Details}
Our whole project is implemented in PyTorch. 
We use the Adam~\cite{adam} optimizer with a learning rate set to \(1\times10^{-4}\). The batch size is 1, and training is iterated for \(500\) epochs. The entire experiment, including training and testing, is conducted on a computer with one NVIDIA RTX 3090 GPU. \(\lambda\) in \eqref{total} is set to 1.

\subsubsection{Evaluating Metrics}
The registration performance is evaluated by the following evaluation metrics:

1) Dice Score(DSC). Dice (Dice Similarity Coefficient, DSC) can determine the degree of overlap in images~\cite{DSC}, that is, the volume overlap between quantified structures. Its value range is [0,1]. A Dice value of 1 for a completely overlapping region. The Dice value explicitly measures the overlap between two regions and thus reflects the quality of the registration. Considering the multiple anatomical structures that have been annotated, we calculated the Dice score for each structure and averaged it. A higher Dice Score indicates more accurate information about the deformation field.
\begin{equation}
    DSC=\frac{2\left |X+Y \right |}{ \left| X\right |+\left| Y\right |}
\end{equation}
X and Y are the binarization results of the two images, $\left | X\cap Y \right | $ denoting the number of elements common to the two images, $\left | X \right | $ and $\left | Y \right | $ denote the total number of elements in the two images.

2) Jacobian Determinant. To quantify the regularity of the deformation fields, we reported the percentages of non-positive values in the determinant of the Jacobian matrix(\(J_{\phi  }(p) = \bigtriangledown_{\phi  }(p)\)) on the deformation field, calculate the count of non-background voxels for which \(|J_{\phi  }| < 0\), indicating regions where the deformation deviates from being diffeomorphic~\cite{JacobianDeterminant}.

\subsection{Ablation Studies}
In this subsection, we give an in-depth analysis of our framework. 
For fair comparisons, all our experiments are conducted on LPBA40. 

\subsubsection{Impact of Pretraining Dec1} 
In our framework, the incorporation of a second-level decoder (Dec2) can be appended to any pre-trained U-Net-like architectures with only one encoder and one decoder (Dec1). Besides, it can also be treated as a whole and trained from scratch. To investigate the impact, we further conduct experiments on optimization strategies. Specifically, we first pre-train the encoder and the first decoder (Dec1) before the second level decoder (Dec2) is added. We then introduce Dec2 for overall training. Table \ref{table-pretrain} compares the experimental results on optimization strategies with or without the pertaining process.
\begin{table}[htbp]
\caption{Results of ablation experiments for whether or not to load pre-trained models.}
\begin{tabular}{cccc}
\toprule 
Model& Pre-trained models & DSC&    $\%|J_{\phi }| < 0$\\ 
\midrule 
\multirow{2}*{U-UNet} 
      &  $\times$   & 0.665    &  0.402   \\
      &  \checkmark & 0.667    &  0.388   \\
\multirow{2}*{U-UNet++} 
        & $\times$  &0.672    &  0.312   \\
        & \checkmark  &0.675    &  0.294   \\
\multirow{2}*{U-UNet3+} 
        & $\times$  &0.672	&  0.266      \\
        & \checkmark   &0.675	&  0.258      \\
\multirow{2}*{U-DenseUnet} 
        & $\times$ &0.677  &  0.196   \\
        & \checkmark  &0.679  &  0.178   \\
\bottomrule 
\end{tabular}
\label{table-pretrain}
\end{table}
As shown in Table~\ref{table-pretrain}, it can be seen that pre-training the UNet network in advance can improve the model performance steadily. Such a strategy has consistently given better results on four types of decoders covering U-UNet, U-UNet++, U-UNet3+, and U-DenseUNet. Specifically, the average Dice scores have been improved by 0.2\%, 0.3\%, 0.3\%, and 0.2\% in respectively, and there is also a reduction in the percentage of voxels with a non-positive Jacobian determinant.

\subsubsection{Impact of Skip-Connections between Dec2 and Encoder} 
The second-level decoder (Dec2) not only fuses the features from the corresponding layer of Decoder 1 (Dec1) but also introduces the features of the corresponding encoder on top of it to derive more expressive features. This method of combining shallow features covers more comprehensive feature information and improves the decoding quality while avoiding losing valuable information. To verify the effectiveness of this design, we also conduct ablation experiments on the benefits of adding skip connections to encoder features in the second-level decoder. Table~\ref{table-encoder} shows the comparison of the experimental results. As shown in Table ~\ref{table-encoder}, we can observe that the average Dice scores are improved with the encoder skip connection strategy on four different network architectures. The average Dice scores of U-UNet are improved by 0.4\%, 0.2\% for U-UNet++, 0.3\% for U-UNet3+, and 0.2\% for U-DenseUNet, demonstrating the effectiveness of the encoder skip connection strategy in improving the image registration accuracy, and that this strategy exhibits some versatility across different network architectures.
\begin{table}[htbp]
\setlength{\belowcaptionskip}{-0.2cm}%
\caption{Results of ablation experiment for the inclusion or exclusion of encoder features.}
\begin{tabular}{cccc}
\toprule 
Model& Encoder Features & DSC&    $\%|J_{\phi  }| < 0$\\ 
\midrule 
\multirow{2}*{U-UNet} 
      &  $\times$   & 0.667    &  0.388   \\
      &  \checkmark & 0.671    &  0.374   \\
\multirow{2}*{U-UNet++} 
        & $\times$ &0.675    &  0.294   \\
        & \checkmark  &0.677    &  0.242   \\
\multirow{2}*{U-UNet3+} 
        & $\times$ &0.675	&  0.258      \\
        & \checkmark &0.678	&  0.233     \\
\multirow{2}*{U-DenseUNet} 
        &$\times$ &0.679  &  0.178   \\
        &\checkmark  &0.681  &  0.164   \\
\bottomrule 
\end{tabular}
\label{table-encoder}
\end{table}

\subsubsection{Impact of Adopting More Level Decoders} We further study the effect of using more decoders. The results of using more levels of decoders are shown in Table~\ref{table-decoder-num}. Note that all decoders here are UNet's decoder. It can be noted that with the increase in the number of decoders, the effect of registration is gradually improved, but at the same time, the increase in the number of decoders also leads to a significant increase in the number of parameters of the model. Therefore, considering the trade-off between the registration accuracy and computational complexity, we adopt a two-level decoder configuration as our default configuration in all our rest experiments.  
\begin{table}[htbp]
\caption{Results of ablation experiment for different level of decoders.}
\begin{tabular}{cccc}
\toprule 
Level of decoder& DSC&$\%|J_{\phi  }| < 0$&Parameters(M)\\ 
\midrule 
 1     &  0.657 & 0.384    &  0.359   \\
 2     &  0.671 & 0.374    &  0.652   \\
 3     &  0.675 & 0.243    &  1.029   \\
 4     &  0.679 & 0.274    &  1.491   \\
\bottomrule 
\end{tabular}
\label{table-decoder-num}
\end{table}

\subsubsection{Impact of different Dec2 Structure}
We further compare the structure using different Dec2 with several popular image registration methods, including Affine transformation, three traditional registration methods SyN, UtilzReg and NiftyReg, and one deep learning-based method VoxelMorph. 
U-UNet, U-UNet++, U-UNet3+, U-DenseUnet are variants of Tetrahedron-Net using UNet as a baseline with different Dec2. These models are loaded with pre-trained Enc and Dec1 and fused encoder features through skip connections.

The results in table~\ref{table-udec} are obtained by training on the LPBA40 dataset. It can be seen that the network using the two-level decoder structure outperforms the original registration network, improving on both evaluation metrics. Encouragingly, with U-UNet, U-UNet++, U-UNet3+, and U-DenseUNet as the structure, Tetrahedron-Net outperforms the baselines by 1.3\%, 1.9\%, 2.0\% and 2.3\%, respectively. Among them, the model using the DenseUnet decoder as Dec2 achieves the best results, obtaining the highest average Dice score of 0.681, which improves by 15.0\%, 1.6\%, and 2.3\% over Affine, UtilzReg, and VoxelMorph, respectively.
\begin{table}[htbp]
\setlength{\belowcaptionskip}{-0.2cm}%
\caption{Quantitative Comparison of Results of Different Registration Methods on LPBA40 dataset.}
\begin{tabular}{cccc}
\toprule 
Methods&  DSC $\uparrow$ &  $\%|J_{\phi  }| < 0 \downarrow$ &Parameters(M)\\ 
\midrule 
Affine only&0.531&	 -    & 	-\\
UtilzReg&0.665&	 -    & 	-\\
NiftyReg &	0.691&	1.13e-3 &   -      \\
ANTs SyN&	0.703&  1.18e-4 &   -      \\
VoxelMorph&	0.658&	0.384   &	0.359        \\
U-UNet&	   0.671    &  0.374   &	0.652        \\
U-UNet++&  0.677    &  0.242   &	1.349        \\
U-UNet3+&  0.678	&  0.233   &	0.723        \\
U-DenseUnet& 0.681  &  0.164   &	1.439        \\
\bottomrule 
\end{tabular}
\label{table-udec}
\end{table}

\subsubsection{Visualization} 
\begin{figure}[htbp]
    \centering
    \begin{minipage}{0.49\linewidth}
        \centering
        \subfloat[Training loss]{\includegraphics[width=1.0\linewidth]{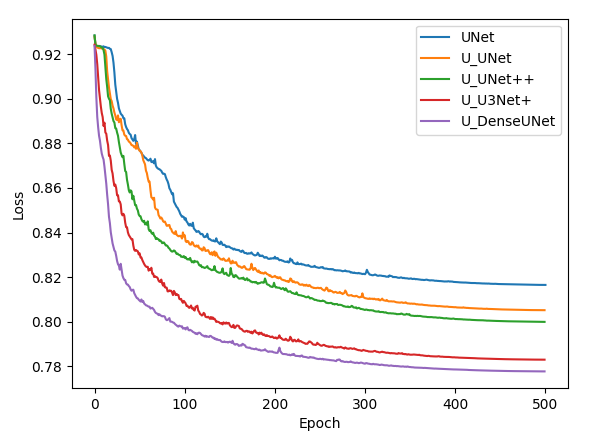}}
    \end{minipage}
    \hfill
    \begin{minipage}{0.49\linewidth}
        \centering
        \subfloat[Dice scores]{\includegraphics[width=1.0\linewidth]{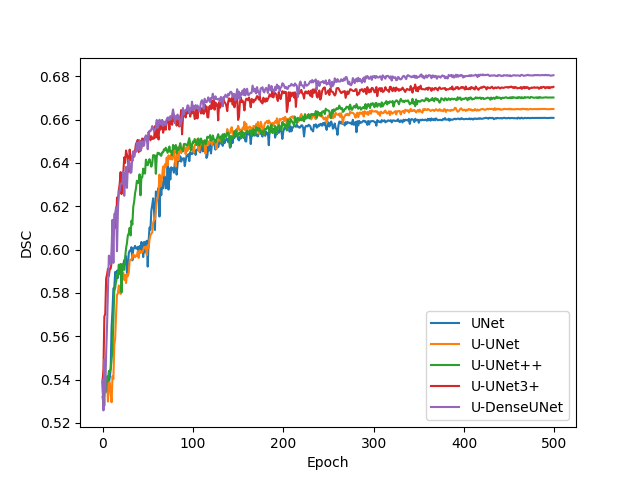}}
    \end{minipage}
    \caption{Training Loss and Validation Dice scores for LPBA40 during training.}
    \label{fig-lpba40-lossanddsc}
\end{figure}

The visualization in Figure~\ref{fig-lpba40-lossanddsc} clearly shows the trend of the loss function and validation dice scores when different Dec2 are used during the training process. When using Decoder 2 (Dec2), the four-way methods have much faster convergence speed and smoothing of the loss curves than using only one decoder. Moreover, when DenseUNet is selected as the second-level decoder, the loss function used decreases faster and converges faster, which is a better performance compared to the other three decoders. In addition, the curve performance of DenseUNet is smoother, which indicates that its training process is more stable and less prone to large fluctuations or oscillations. This result fully proves the advantage of DenseUNet when used as Dec2.

\begin{figure}[htbp]
\setlength{\belowcaptionskip}{-0.2cm}%
\centering
\includegraphics[width=1\linewidth]{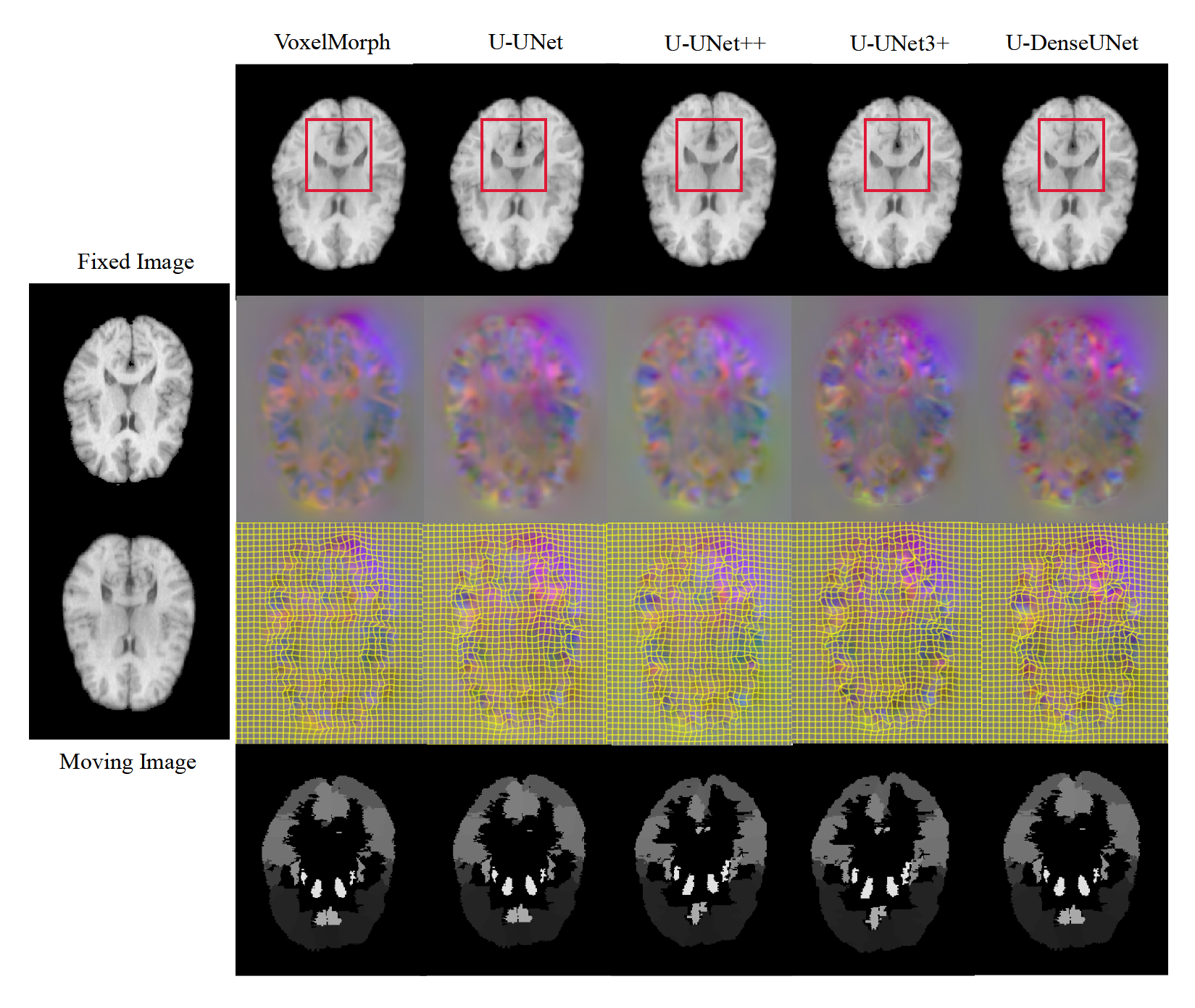}
\caption{Visualisation of results of image registration using different Dec2 on LPBA40.}
\label{figrure:LPBA40 result}
\end{figure}

Figure~\ref{figrure:LPBA40 result} shows the visualization of the results of the different registration methods on LPBA40.  On the left side, it shows the input image pairs, fixed image ($f$), and moving images ($m$). The right side exhibits, from top to bottom, the deformed images, deformation fields images, deformation fields grid images and label images of the deformed images. It can be seen that the addition of Dec2 results in improved accuracy in the detailed parts of the image, especially the parts framed by the red box, and a smoother deformation field.

\subsection{Comparison with State-of-the-arts}

\begin{figure*}[h!tb]
\begin{center}
\centerline{\includegraphics[width=1\linewidth]{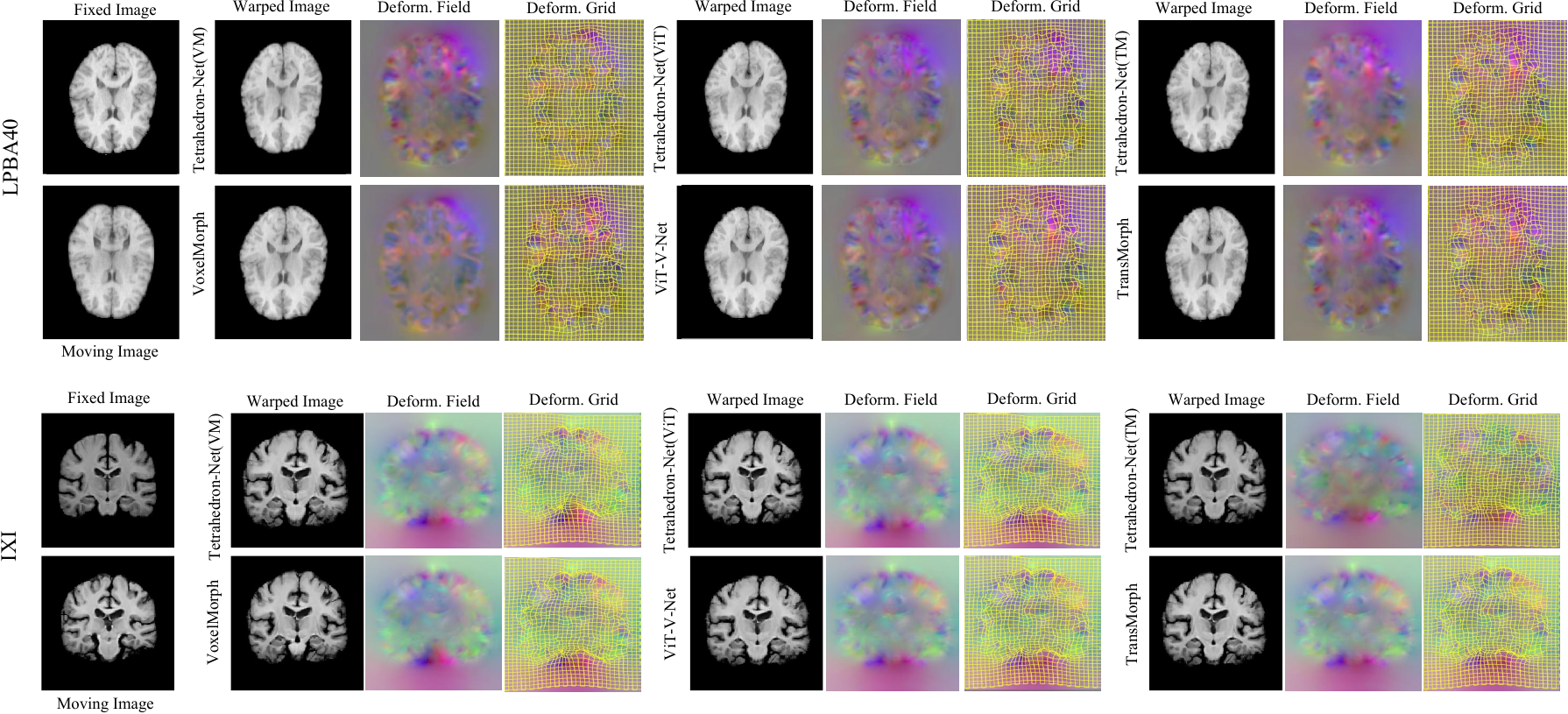}}
\caption{Visualization of registration results on datasets LPBA40, IXI. These are the images from baseline VoxelMorph, ViT-V-Net, TransMorph (row 2), and their respective model trained with Dec2 (row 1).}
\label{fig-model-Dec2}
\end{center}
\end{figure*}

Table~\ref{tabel-model-Dec2} shows the results of models incorporating Dec2 in different registration methods on LPBA40, IXI, and OASIS. We use DenseUNet as the Dec2, which gave the best results in the ablation studies.

\subsubsection{IXI} On the IXI dataset, the model with the addition of Dec2 outperforms the previous state-of-the-art TransMorph-bspl method, with a +0.4\% improvement in DSC, reduced the percentages of non-diffeomorphic voxels ($\%|J_{\phi  }| < 0$). In addition, better results are obtained on other deep learning-based methods, with a +2.0 \% improvement on DSC for VoxelMorph, and a + 1.1\% improvement on DSC for ViT-V-Net, further validating the effectiveness and generality of our method.

\subsubsection{LPBA40} For the LPBA40 dataset, it has only 40 volumes, and on such a small dataset, our method can still greatly improve the learning ability of the model and achieve better results, with a +2.3\% improvement on DSC for VoxelMorph, a +1.0\% improvement on DSC for ViT-V-Net, and a +0.6\% improvement on DSC for TransMorph.

\subsubsection{OASIS} On the OASIS dataset, the incorporation of Dec2 also gives an outstanding performance. Specifically, it has considerably increased DSC over the compared learning-based models. Compared with ViT-V-Net, we outperform it by +1.3\% in DSC. Compared with TransMorph, we increase the DSC by +1.0\%. It once again proves the strong capacity of our framework for MIR. 
\begin{table}[h!tb]
\caption{Registration experiment results of Tetrahedron-Net using different networks as baseline on LPBA40, IXI, and OASIS. VM, ViT, TM, TM-bspl represent VoxelMorph,  ViT-V-Net, TransMorph, and TransMorph-bspl.}
\resizebox{0.5\textwidth}{!}{ 
    \setlength{\tabcolsep}{1mm}{
        \begin{tabular}{cccc}
        \toprule 
        Datasets&Methods&  DSC $\uparrow$ &  $\%|J_{\phi  }| < 0 \downarrow$ \\ 
        \midrule 
        \multirow{8}*{LPBA40} 
                & NiftyReg~\cite{NiftyReg} &	0.691& 1.13e-3 \\
        ~       & ANTs SyN~\cite{syn}      &	0.703& 1.18e-4 \\
        ~       & VoxelMorph~\cite{Voxelmorph}& 0.658&  0.288 \\
        ~       & ViT-V-Net~\cite{chen2021vit}&	    0.663& 	0.390 \\
        ~       & TransMorph~\cite{transmorph}&	0.678&	0.438 \\
                & \textbf{Tetrahedron-Net(VM)}&\textbf{0.681}&\textbf{0.164} \\
                & \textbf{Tetrahedron-Net(ViT)}&\textbf{0.673}&\textbf{0.363} \\
                & \textbf{Tetrahedron-Net(TM)}&\textbf{0.684}&\textbf{0.285}\\
        \midrule 
        \multirow{10}*{IXI} 
               & NiftyReg~\cite{NiftyReg} &	  0.585&0.029\\
               & ANTs SyN~\cite{syn}      &      0.647&1.96e-6 \\
               & VoxelMorph~\cite{Voxelmorph}&   0.729&  1.590 \\
               & ViT-V-Net~\cite{chen2021vit}&	  0.734&  1.609 \\
               & TransMorph-bspl~\cite{transmorph}&0.761&<0.0001\\
               & TransMorph~\cite{transmorph}&	  0.753&  1.579 \\
               & \textbf{Tetrahedron-Net(VM)}&\textbf{0.749}&\textbf{1.326}\\
               & \textbf{Tetrahedron-Net(ViT)}&\textbf{0.745}&\textbf{1.535} \\
               & \textbf{Tetrahedron-Net(TM)}&\textbf{0.757}&\textbf{1.186} \\   
               & \textbf{Tetrahedron-Net(TM-bspl)}&\textbf{0.765}&\textbf{<0.0001}\\
        \midrule 
        \multirow{6}*{OASIS} 
               & NiftyReg~\cite{NiftyReg} &	0.762&0.011\\
               & ANTs SyN~\cite{syn}      &    0.769&1.58e-4\\
               & ViT-V-Net~\cite{chen2021vit} &	0.794& 0.887 \\
               & TransMorph~\cite{transmorph}&	0.818& 0.765 \\
               & \textbf{Tetrahedron-Net(ViT)}  &\textbf{0.807}&\textbf{0.876} \\
               & \textbf{Tetrahedron-Net(TM)}&	\textbf{0.828}&\textbf{0.745}\\
        \bottomrule
        \end{tabular}
    }
}
\label{tabel-model-Dec2}
\end{table}
\subsubsection{Visualization} In Figure ~\ref{fig-model-Dec2}, we show visualized images of the registration results on different datasets. Comparing the baseline model incorporating Dec2, it can be observed that the results obtained from the network with Dec2 are better at handling complex scenarios involving intricate details and have smoother deformation fields. 

\begin{figure}[htbp]
    \centering
    \begin{minipage}{0.49\linewidth}
        \centering
        \subfloat[LPBA40]{\includegraphics[width=1.0\linewidth]{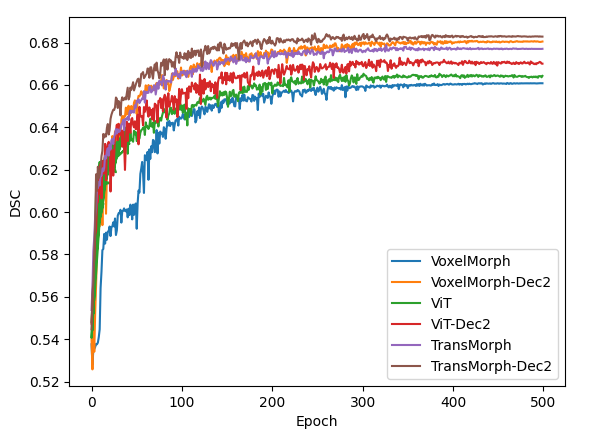}}
    \end{minipage}
    \hfill
    \begin{minipage}{0.49\linewidth}
        \centering
        \subfloat[IXI]{\includegraphics[width=1.0\linewidth]{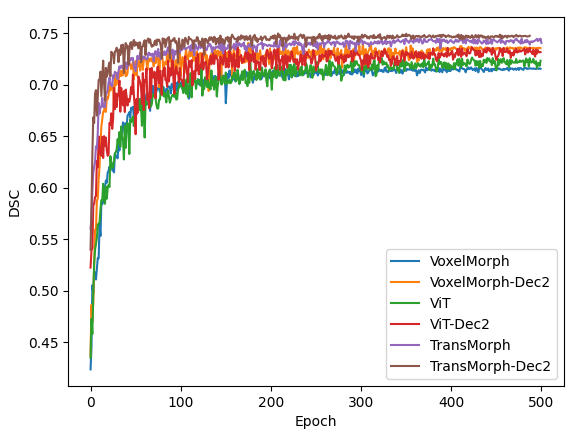}}
    \end{minipage}
    \caption{Validation Dice scores on LPBA40 and IXI during training with Dec2 on different registration models.}
    \label{fig-dsc-lpba40ixi}
\end{figure}

Figure~\ref{fig-dsc-lpba40ixi} plots DSC trends of deep learning models including VoxelMorph, ViT-V-Net, and TransMorph on LPBA40 and IXI. It can be seen that with the addition of Dec2 they demonstrate a faster convergence speed. Besides, Dec2 achieves the better results when combined with the convolution-only VoxelMorph model, giving us the best Dice scores and showing smoother curves.


\section{Conclusions}
In this paper, we have presented a simple yet effective framework named Tetrahedron-Net for unsupervised 3D medical image registration. The architecture of this framework incorporates one encoder and a two-level decoder. This design allows the features obtained by the encoder to be decoded twice to estimate the deformation field more accurately. 
Extensive experiments has been conducted on LPBA40, IXI, and OAISIS, showing that the proposed Tetrahedron-Net can outperform the state-of-the-art methods by a large margin.

\noindent\textbf{Limitations and future work.} our work also has some limitations. 
Its effectiveness is validated only on medical image registration. 
However, it is indeed a general backbone and can be applied in many dense prediction tasks in medical and computer vision fields.  
In the future, we plan to apply it in more tasks such as medical image segmentation and object detection to study its generality.

\printcredits

\bibliographystyle{cas-model2-names}

\bibliography{Tetrahedron-Net}

\begin{thebibliography}{50}
\expandafter\ifx\csname natexlab\endcsname\relax\def\natexlab#1{#1}\fi
\providecommand{\url}[1]{\texttt{#1}}
\providecommand{\href}[2]{#2}
\providecommand{\path}[1]{#1}
\providecommand{\DOIprefix}{doi:}
\providecommand{\ArXivprefix}{arXiv:}
\providecommand{\URLprefix}{URL: }
\providecommand{\Pubmedprefix}{pmid:}
\providecommand{\doi}[1]{\href{http://dx.doi.org/#1}{\path{#1}}}
\providecommand{\Pubmed}[1]{\href{pmid:#1}{\path{#1}}}
\providecommand{\bibinfo}[2]{#2}
\ifx\xfnm\relax \def\xfnm[#1]{\unskip,\space#1}\fi
\bibitem[{Alam et~al.(2018)Alam, Rahman, Ullah and Gulati}]{imageguided}
\bibinfo{author}{Alam, F.}, \bibinfo{author}{Rahman, S.U.}, \bibinfo{author}{Ullah, S.}, \bibinfo{author}{Gulati, K.}, \bibinfo{year}{2018}.
\newblock \bibinfo{title}{Medical image registration in image guided surgery: Issues, challenges and research opportunities}.
\newblock \bibinfo{journal}{Biocybernetics and Biomedical Engineering} \bibinfo{volume}{38}, \bibinfo{pages}{71--89}.
\bibitem[{Ashburner(2007)}]{JacobianDeterminant}
\bibinfo{author}{Ashburner, J.}, \bibinfo{year}{2007}.
\newblock \bibinfo{title}{A fast diffeomorphic image registration algorithm}.
\newblock \bibinfo{journal}{Neuroimage} \bibinfo{volume}{38}, \bibinfo{pages}{95--113}.
\bibitem[{Avants et~al.(2008)Avants, Epstein, Grossman and Gee}]{syn}
\bibinfo{author}{Avants, B.B.}, \bibinfo{author}{Epstein, C.L.}, \bibinfo{author}{Grossman, M.}, \bibinfo{author}{Gee, J.C.}, \bibinfo{year}{2008}.
\newblock \bibinfo{title}{Symmetric diffeomorphic image registration with cross-correlation: evaluating automated labeling of elderly and neurodegenerative brain}.
\newblock \bibinfo{journal}{Medical image analysis} \bibinfo{volume}{12}, \bibinfo{pages}{26--41}.
\bibitem[{Azad et~al.(2023)Azad, Kazerouni, Heidari, Aghdam, Molaei, Jia, Jose, Roy and Merhof}]{vitreview}
\bibinfo{author}{Azad, R.}, \bibinfo{author}{Kazerouni, A.}, \bibinfo{author}{Heidari, M.}, \bibinfo{author}{Aghdam, E.K.}, \bibinfo{author}{Molaei, A.}, \bibinfo{author}{Jia, Y.}, \bibinfo{author}{Jose, A.}, \bibinfo{author}{Roy, R.}, \bibinfo{author}{Merhof, D.}, \bibinfo{year}{2023}.
\newblock \bibinfo{title}{Advances in medical image analysis with vision transformers: a comprehensive review}.
\newblock \bibinfo{journal}{Medical Image Analysis} , \bibinfo{pages}{103000}.
\bibitem[{Bajcsy and Kova{\v{c}}i{\v{c}}(1989)}]{DSC}
\bibinfo{author}{Bajcsy, R.}, \bibinfo{author}{Kova{\v{c}}i{\v{c}}, S.}, \bibinfo{year}{1989}.
\newblock \bibinfo{title}{Multiresolution elastic matching}.
\newblock \bibinfo{journal}{Computer vision, graphics, and image processing} \bibinfo{volume}{46}, \bibinfo{pages}{1--21}.
\bibitem[{Balakrishnan et~al.(2019)Balakrishnan, Zhao, Sabuncu, Guttag and Dalca}]{Voxelmorph}
\bibinfo{author}{Balakrishnan, G.}, \bibinfo{author}{Zhao, A.}, \bibinfo{author}{Sabuncu, M.R.}, \bibinfo{author}{Guttag, J.}, \bibinfo{author}{Dalca, A.V.}, \bibinfo{year}{2019}.
\newblock \bibinfo{title}{Voxelmorph: a learning framework for deformable medical image registration}.
\newblock \bibinfo{journal}{IEEE Transactions on medical imaging} \bibinfo{volume}{38}, \bibinfo{pages}{1788--1800}.
\bibitem[{Chen et~al.(2022)Chen, Frey, He, Segars, Li and Du}]{transmorph}
\bibinfo{author}{Chen, J.}, \bibinfo{author}{Frey, E.C.}, \bibinfo{author}{He, Y.}, \bibinfo{author}{Segars, W.P.}, \bibinfo{author}{Li, Y.}, \bibinfo{author}{Du, Y.}, \bibinfo{year}{2022}.
\newblock \bibinfo{title}{Transmorph: Transformer for unsupervised medical image registration}.
\newblock \bibinfo{journal}{Medical image analysis} \bibinfo{volume}{82}, \bibinfo{pages}{102615}.
\bibitem[{Chen et~al.(2021a)Chen, He, Frey, Li and Du}]{chen2021vit}
\bibinfo{author}{Chen, J.}, \bibinfo{author}{He, Y.}, \bibinfo{author}{Frey, E.C.}, \bibinfo{author}{Li, Y.}, \bibinfo{author}{Du, Y.}, \bibinfo{year}{2021}a.
\newblock \bibinfo{title}{Vit-v-net: Vision transformer for unsupervised volumetric medical image registration}.
\newblock \bibinfo{journal}{arXiv preprint arXiv:2104.06468} .
\bibitem[{Chen et~al.(2021b)Chen, Diaz-Pinto, Ravikumar and Frangi}]{review2021}
\bibinfo{author}{Chen, X.}, \bibinfo{author}{Diaz-Pinto, A.}, \bibinfo{author}{Ravikumar, N.}, \bibinfo{author}{Frangi, A.F.}, \bibinfo{year}{2021}b.
\newblock \bibinfo{title}{Deep learning in medical image registration}.
\newblock \bibinfo{journal}{Progress in Biomedical Engineering} \bibinfo{volume}{3}, \bibinfo{pages}{012003}.
\bibitem[{Christensen and Johnson(2001)}]{elastic}
\bibinfo{author}{Christensen, G.E.}, \bibinfo{author}{Johnson, H.J.}, \bibinfo{year}{2001}.
\newblock \bibinfo{title}{Consistent image registration}.
\newblock \bibinfo{journal}{IEEE transactions on medical imaging} \bibinfo{volume}{20}, \bibinfo{pages}{568--582}.
\bibitem[{Delmon et~al.(2013)Delmon, Rit, Pinho and Sarrut}]{b-spline}
\bibinfo{author}{Delmon, V.}, \bibinfo{author}{Rit, S.}, \bibinfo{author}{Pinho, R.}, \bibinfo{author}{Sarrut, D.}, \bibinfo{year}{2013}.
\newblock \bibinfo{title}{Registration of sliding objects using direction dependent b-splines decomposition*}.
\newblock \bibinfo{journal}{Physics in Medicine \& Biology} \bibinfo{volume}{58}, \bibinfo{pages}{1303}.
\bibitem[{Dosovitskiy(2020)}]{dosovitskiy2020image}
\bibinfo{author}{Dosovitskiy, A.}, \bibinfo{year}{2020}.
\newblock \bibinfo{title}{An image is worth 16x16 words: Transformers for image recognition at scale}.
\newblock \bibinfo{journal}{arXiv preprint arXiv:2010.11929} .
\bibitem[{Estienne et~al.(2019)Estienne, Vakalopoulou, Christodoulidis, Battistela, Lerousseau, Carre, Klausner, Sun, Robert, Mougiakakou et~al.}]{u-ReSNet}
\bibinfo{author}{Estienne, T.}, \bibinfo{author}{Vakalopoulou, M.}, \bibinfo{author}{Christodoulidis, S.}, \bibinfo{author}{Battistela, E.}, \bibinfo{author}{Lerousseau, M.}, \bibinfo{author}{Carre, A.}, \bibinfo{author}{Klausner, G.}, \bibinfo{author}{Sun, R.}, \bibinfo{author}{Robert, C.}, \bibinfo{author}{Mougiakakou, S.}, et~al., \bibinfo{year}{2019}.
\newblock \bibinfo{title}{U-resnet: Ultimate coupling of registration and segmentation with deep nets}, in: \bibinfo{booktitle}{Medical Image Computing and Computer Assisted Intervention--MICCAI 2019: 22nd International Conference, Shenzhen, China, October 13--17, 2019, Proceedings, Part III 22}, pp. \bibinfo{pages}{310--319}.
\bibitem[{Fischl(2012)}]{FreeSurfer}
\bibinfo{author}{Fischl, B.}, \bibinfo{year}{2012}.
\newblock \bibinfo{title}{Freesurfer}.
\newblock \bibinfo{journal}{Neuroimage} \bibinfo{volume}{62}, \bibinfo{pages}{774--781}.
\bibitem[{Hammoudeh and Dupont(2023)}]{review2023}
\bibinfo{author}{Hammoudeh, A.}, \bibinfo{author}{Dupont, S.}, \bibinfo{year}{2023}.
\newblock \bibinfo{title}{Deep learning in medical image registration: introduction and survey}.
\newblock \bibinfo{journal}{arXiv preprint arXiv:2309.00727} .
\bibitem[{He et~al.(2023)He, Gan, Li, Rekik, Yin, Ji, Gao, Wang, Zhang and Shen}]{transformersreview}
\bibinfo{author}{He, K.}, \bibinfo{author}{Gan, C.}, \bibinfo{author}{Li, Z.}, \bibinfo{author}{Rekik, I.}, \bibinfo{author}{Yin, Z.}, \bibinfo{author}{Ji, W.}, \bibinfo{author}{Gao, Y.}, \bibinfo{author}{Wang, Q.}, \bibinfo{author}{Zhang, J.}, \bibinfo{author}{Shen, D.}, \bibinfo{year}{2023}.
\newblock \bibinfo{title}{Transformers in medical image analysis}.
\newblock \bibinfo{journal}{Intelligent Medicine} \bibinfo{volume}{3}, \bibinfo{pages}{59--78}.
\bibitem[{He et~al.(2016)He, Zhang, Ren and Sun}]{resnet}
\bibinfo{author}{He, K.}, \bibinfo{author}{Zhang, X.}, \bibinfo{author}{Ren, S.}, \bibinfo{author}{Sun, J.}, \bibinfo{year}{2016}.
\newblock \bibinfo{title}{Deep residual learning for image recognition}, in: \bibinfo{booktitle}{Proceedings of the IEEE conference on computer vision and pattern recognition}, pp. \bibinfo{pages}{770--778}.
\bibitem[{Hu et~al.(2023)Hu, Zhang, Matkovic, Liu and Yang}]{2023reinforcement}
\bibinfo{author}{Hu, M.}, \bibinfo{author}{Zhang, J.}, \bibinfo{author}{Matkovic, L.}, \bibinfo{author}{Liu, T.}, \bibinfo{author}{Yang, X.}, \bibinfo{year}{2023}.
\newblock \bibinfo{title}{Reinforcement learning in medical image analysis: Concepts, applications, challenges, and future directions}.
\newblock \bibinfo{journal}{Journal of Applied Clinical Medical Physics} \bibinfo{volume}{24}, \bibinfo{pages}{e13898}.
\bibitem[{Huang et~al.(2017)Huang, Liu, Van Der~Maaten and Weinberger}]{huang2017densely}
\bibinfo{author}{Huang, G.}, \bibinfo{author}{Liu, Z.}, \bibinfo{author}{Van Der~Maaten, L.}, \bibinfo{author}{Weinberger, K.Q.}, \bibinfo{year}{2017}.
\newblock \bibinfo{title}{Densely connected convolutional networks}, in: \bibinfo{booktitle}{Proceedings of the IEEE conference on computer vision and pattern recognition}, pp. \bibinfo{pages}{4700--4708}.
\bibitem[{Huang et~al.(2020)Huang, Lin, Tong, Hu, Zhang, Iwamoto, Han, Chen and Wu}]{unet3+}
\bibinfo{author}{Huang, H.}, \bibinfo{author}{Lin, L.}, \bibinfo{author}{Tong, R.}, \bibinfo{author}{Hu, H.}, \bibinfo{author}{Zhang, Q.}, \bibinfo{author}{Iwamoto, Y.}, \bibinfo{author}{Han, X.}, \bibinfo{author}{Chen, Y.W.}, \bibinfo{author}{Wu, J.}, \bibinfo{year}{2020}.
\newblock \bibinfo{title}{Unet 3+: A full-scale connected unet for medical image segmentation}, in: \bibinfo{booktitle}{ICASSP 2020-2020 IEEE international conference on acoustics, speech and signal processing (ICASSP)}, pp. \bibinfo{pages}{1055--1059}.
\bibitem[{Huang et~al.(2022)Huang, Chen, Chen, Chen and Wan}]{huang2022tdd}
\bibinfo{author}{Huang, X.}, \bibinfo{author}{Chen, J.}, \bibinfo{author}{Chen, M.}, \bibinfo{author}{Chen, L.}, \bibinfo{author}{Wan, Y.}, \bibinfo{year}{2022}.
\newblock \bibinfo{title}{Tdd-unet: Transformer with double decoder unet for covid-19 lesions segmentation}.
\newblock \bibinfo{journal}{Computers in Biology and Medicine} \bibinfo{volume}{151}, \bibinfo{pages}{106306}.
\bibitem[{Jaderberg et~al.(2015)Jaderberg, Simonyan, Zisserman et~al.}]{stn}
\bibinfo{author}{Jaderberg, M.}, \bibinfo{author}{Simonyan, K.}, \bibinfo{author}{Zisserman, A.}, et~al., \bibinfo{year}{2015}.
\newblock \bibinfo{title}{Spatial transformer networks}.
\newblock \bibinfo{journal}{Advances in neural information processing systems} \bibinfo{volume}{28}.
\bibitem[{Jia et~al.(2022)Jia, Bartlett, Zhang, Lu, Qiu and Duan}]{Jia2022UNetVT}
\bibinfo{author}{Jia, X.}, \bibinfo{author}{Bartlett, J.}, \bibinfo{author}{Zhang, T.}, \bibinfo{author}{Lu, W.}, \bibinfo{author}{Qiu, Z.}, \bibinfo{author}{Duan, J.}, \bibinfo{year}{2022}.
\newblock \bibinfo{title}{U-net vs transformer: Is u-net outdated in medical image registration?}, in: \bibinfo{booktitle}{MLMI@MICCAI}.
\bibitem[{Kim et~al.(2020)Kim, Kim, Park, Kim, Lee and Ye}]{CycleMorph}
\bibinfo{author}{Kim, B.}, \bibinfo{author}{Kim, D.H.}, \bibinfo{author}{Park, S.H.}, \bibinfo{author}{Kim, J.}, \bibinfo{author}{Lee, J.G.}, \bibinfo{author}{Ye, J.C.}, \bibinfo{year}{2020}.
\newblock \bibinfo{title}{Cyclemorph: Cycle consistent unsupervised deformable image registration}.
\newblock \bibinfo{journal}{Medical image analysis} \bibinfo{volume}{71}, \bibinfo{pages}{102036}.
\bibitem[{Kingma and Ba(2014)}]{adam}
\bibinfo{author}{Kingma, D.P.}, \bibinfo{author}{Ba, J.}, \bibinfo{year}{2014}.
\newblock \bibinfo{title}{Adam: A method for stochastic optimization}.
\newblock \bibinfo{journal}{arXiv preprint arXiv:1412.6980} .
\bibitem[{Kuang and Schmah(2018)}]{faim}
\bibinfo{author}{Kuang, D.}, \bibinfo{author}{Schmah, T.}, \bibinfo{year}{2018}.
\newblock \bibinfo{title}{Faim - a convnet method for unsupervised 3d medical image registration}.
\newblock \bibinfo{journal}{ArXiv} \bibinfo{volume}{abs/1811.09243}.
\bibitem[{Li et~al.(2023)Li, Chen, Tang, Wang, Landman and Zhou}]{Transforming}
\bibinfo{author}{Li, J.}, \bibinfo{author}{Chen, J.}, \bibinfo{author}{Tang, Y.}, \bibinfo{author}{Wang, C.}, \bibinfo{author}{Landman, B.A.}, \bibinfo{author}{Zhou, S.K.}, \bibinfo{year}{2023}.
\newblock \bibinfo{title}{Transforming medical imaging with transformers? a comparative review of key properties, current progresses, and future perspectives}.
\newblock \bibinfo{journal}{Medical Image Analysis} \bibinfo{volume}{85}, \bibinfo{pages}{102762}.
\newblock \DOIprefix\doi{https://doi.org/10.1016/j.media.2023.102762}.
\bibitem[{Liu et~al.(2021)Liu, Lin, Cao, Hu, Wei, Zhang, Lin and Guo}]{swintransformer}
\bibinfo{author}{Liu, Z.}, \bibinfo{author}{Lin, Y.}, \bibinfo{author}{Cao, Y.}, \bibinfo{author}{Hu, H.}, \bibinfo{author}{Wei, Y.}, \bibinfo{author}{Zhang, Z.}, \bibinfo{author}{Lin, S.}, \bibinfo{author}{Guo, B.}, \bibinfo{year}{2021}.
\newblock \bibinfo{title}{Swin transformer: Hierarchical vision transformer using shifted windows}, in: \bibinfo{booktitle}{Proceedings of the IEEE/CVF international conference on computer vision}, pp. \bibinfo{pages}{10012--10022}.
\bibitem[{London(2023)}]{ixi}
\bibinfo{author}{London, I.C.}, \bibinfo{year}{2023}.
\newblock \bibinfo{title}{Information extration from images}.
\newblock \bibinfo{howpublished}{https://brain-development.org/ixi-dataset/}.
\bibitem[{Long et~al.(2015)Long, Shelhamer and Darrell}]{fcn}
\bibinfo{author}{Long, J.}, \bibinfo{author}{Shelhamer, E.}, \bibinfo{author}{Darrell, T.}, \bibinfo{year}{2015}.
\newblock \bibinfo{title}{Fully convolutional networks for semantic segmentation}, in: \bibinfo{booktitle}{Proceedings of the IEEE conference on computer vision and pattern recognition}, pp. \bibinfo{pages}{3431--3440}.
\bibitem[{Marcus et~al.(2014)Marcus, Wang, Parker, Csernansky, Morris and Buckner}]{OASIS}
\bibinfo{author}{Marcus, D.S.}, \bibinfo{author}{Wang, T.H.}, \bibinfo{author}{Parker, J.}, \bibinfo{author}{Csernansky, J.G.}, \bibinfo{author}{Morris, J.C.}, \bibinfo{author}{Buckner, R.L.}, \bibinfo{year}{2014}.
\newblock \bibinfo{title}{Open access series of imaging studies (oasis): Cross-sectional mri data in young, middle aged, nondemented, and demented older adults}.
\newblock \bibinfo{journal}{Journal of Cognitive Neuroscience} \bibinfo{volume}{19}, \bibinfo{pages}{1498--1507}.
\bibitem[{for Medical Image~Computing(2023)}]{NiftyReg}
\bibinfo{author}{for Medical Image~Computing, U.C.}, \bibinfo{year}{2023}.
\newblock \bibinfo{title}{University college london. niftyreg}.
\newblock \bibinfo{howpublished}{http://cmictig.cs.ucl.ac.uk/wiki/index.php/NiftyReg}.
\bibitem[{Oktay et~al.(2018)Oktay, Schlemper, Folgoc, Lee, Heinrich, Misawa, Mori, McDonagh, Hammerla, Kainz et~al.}]{attentionunet}
\bibinfo{author}{Oktay, O.}, \bibinfo{author}{Schlemper, J.}, \bibinfo{author}{Folgoc, L.L.}, \bibinfo{author}{Lee, M.}, \bibinfo{author}{Heinrich, M.}, \bibinfo{author}{Misawa, K.}, \bibinfo{author}{Mori, K.}, \bibinfo{author}{McDonagh, S.}, \bibinfo{author}{Hammerla, N.Y.}, \bibinfo{author}{Kainz, B.}, et~al., \bibinfo{year}{2018}.
\newblock \bibinfo{title}{Attention u-net: Learning where to look for the pancreas}.
\newblock \bibinfo{journal}{arXiv preprint arXiv:1804.03999} .
\bibitem[{Razzak et~al.(2018)Razzak, Naz and Zaib}]{review2018}
\bibinfo{author}{Razzak, M.I.}, \bibinfo{author}{Naz, S.}, \bibinfo{author}{Zaib, A.}, \bibinfo{year}{2018}.
\newblock \bibinfo{title}{Deep learning for medical image processing: Overview, challenges and the future}.
\newblock \bibinfo{journal}{Classification in BioApps: Automation of decision making} , \bibinfo{pages}{323--350}.
\bibitem[{Ronneberger et~al.(2015)Ronneberger, Fischer and Brox}]{unet}
\bibinfo{author}{Ronneberger, O.}, \bibinfo{author}{Fischer, P.}, \bibinfo{author}{Brox, T.}, \bibinfo{year}{2015}.
\newblock \bibinfo{title}{U-net: Convolutional networks for biomedical image segmentation}, in: \bibinfo{booktitle}{International Conference on Medical Image Computing and Computer-Assisted Intervention}.
\bibitem[{Sheikhjafari et~al.(2022)Sheikhjafari, Noga, Punithakumar and Ray}]{2022unsupervised}
\bibinfo{author}{Sheikhjafari, A.}, \bibinfo{author}{Noga, M.}, \bibinfo{author}{Punithakumar, K.}, \bibinfo{author}{Ray, N.}, \bibinfo{year}{2022}.
\newblock \bibinfo{title}{Unsupervised deformable image registration with fully connected generative neural network}, in: \bibinfo{booktitle}{Medical imaging with deep learning}.
\bibitem[{Sokooti et~al.(2017)Sokooti, Vos, Berendsen, Lelieveldt, Isgum and Staring}]{dirnet}
\bibinfo{author}{Sokooti, H.}, \bibinfo{author}{Vos, B.D.}, \bibinfo{author}{Berendsen, F.F.}, \bibinfo{author}{Lelieveldt, B.P.F.}, \bibinfo{author}{Isgum, I.}, \bibinfo{author}{Staring, M.}, \bibinfo{year}{2017}.
\newblock \bibinfo{title}{Nonrigid image registration using multi-scale 3d convolutional neural networks}, in: \bibinfo{booktitle}{International Conference on Medical Image Computing and Computer-Assisted Intervention}.
\bibitem[{of~Neuro Imaging University~of Southern~California(2023)}]{LPBA40}
\bibinfo{author}{of~Neuro Imaging University~of Southern~California, L.}, \bibinfo{year}{2023}.
\newblock \bibinfo{title}{Loni probabilistic brain atlas (lpba40)}.
\newblock \bibinfo{howpublished}{https://loni.usc.edu/research/atlases}.
\bibitem[{Szegedy et~al.(2015)Szegedy, Liu, Jia, Sermanet, Reed, Anguelov, Erhan, Vanhoucke and Rabinovich}]{googlenet}
\bibinfo{author}{Szegedy, C.}, \bibinfo{author}{Liu, W.}, \bibinfo{author}{Jia, Y.}, \bibinfo{author}{Sermanet, P.}, \bibinfo{author}{Reed, S.}, \bibinfo{author}{Anguelov, D.}, \bibinfo{author}{Erhan, D.}, \bibinfo{author}{Vanhoucke, V.}, \bibinfo{author}{Rabinovich, A.}, \bibinfo{year}{2015}.
\newblock \bibinfo{title}{Going deeper with convolutions}, in: \bibinfo{booktitle}{Proceedings of the IEEE conference on computer vision and pattern recognition}, pp. \bibinfo{pages}{1--9}.
\bibitem[{Wang et~al.(2022a)Wang, Cao, Wang and Zaiane}]{UCTransNet}
\bibinfo{author}{Wang, H.}, \bibinfo{author}{Cao, P.}, \bibinfo{author}{Wang, J.}, \bibinfo{author}{Zaiane, O.R.}, \bibinfo{year}{2022}a.
\newblock \bibinfo{title}{Uctransnet: rethinking the skip connections in u-net from a channel-wise perspective with transformer}, in: \bibinfo{booktitle}{Proceedings of the AAAI conference on artificial intelligence}, pp. \bibinfo{pages}{2441--2449}.
\bibitem[{Wang et~al.(2022b)Wang, Qian, Li and Zhang}]{wang2022transformer}
\bibinfo{author}{Wang, Y.}, \bibinfo{author}{Qian, W.}, \bibinfo{author}{Li, M.}, \bibinfo{author}{Zhang, X.}, \bibinfo{year}{2022}b.
\newblock \bibinfo{title}{A transformer-based network for deformable medical image registration}, in: \bibinfo{booktitle}{CAAI International Conference on Artificial Intelligence}, pp. \bibinfo{pages}{502--513}.
\bibitem[{Wu et~al.(2021)Wu, Wu, Jin, Cao and Jin}]{denseunet}
\bibinfo{author}{Wu, Y.}, \bibinfo{author}{Wu, J.}, \bibinfo{author}{Jin, S.}, \bibinfo{author}{Cao, L.}, \bibinfo{author}{Jin, G.}, \bibinfo{year}{2021}.
\newblock \bibinfo{title}{Dense-u-net: dense encoder--decoder network for holographic imaging of 3d particle fields}.
\newblock \bibinfo{journal}{Optics Communications} \bibinfo{volume}{493}, \bibinfo{pages}{126970}.
\bibitem[{Zhang(2018)}]{icnet}
\bibinfo{author}{Zhang, J.}, \bibinfo{year}{2018}.
\newblock \bibinfo{title}{Inverse-consistent deep networks for unsupervised deformable image registration}.
\newblock \bibinfo{journal}{arXiv preprint arXiv:1809.03443} .
\bibitem[{Zhang et~al.(2013)Zhang, Wang, Wang and Feng}]{fluid}
\bibinfo{author}{Zhang, J.}, \bibinfo{author}{Wang, J.}, \bibinfo{author}{Wang, X.}, \bibinfo{author}{Feng, D.}, \bibinfo{year}{2013}.
\newblock \bibinfo{title}{The adaptive fem elastic model for medical image registration}.
\newblock \bibinfo{journal}{Physics in Medicine \& Biology} \bibinfo{volume}{59}, \bibinfo{pages}{97}.
\newblock \URLprefix \url{https://dx.doi.org/10.1088/0031-9155/59/1/97}, \DOIprefix\doi{10.1088/0031-9155/59/1/97}.
\bibitem[{Zhao et~al.(2019a)Zhao, Balakrishnan, Durand, Guttag and Dalca}]{DataAU}
\bibinfo{author}{Zhao, A.}, \bibinfo{author}{Balakrishnan, G.}, \bibinfo{author}{Durand, F.}, \bibinfo{author}{Guttag, J.V.}, \bibinfo{author}{Dalca, A.V.}, \bibinfo{year}{2019}a.
\newblock \bibinfo{title}{Data augmentation using learned transforms for one-shot medical image segmentation}.
\newblock \bibinfo{journal}{ArXiv} \bibinfo{volume}{abs/1902.09383}.
\bibitem[{Zhao et~al.(2019b)Zhao, Dong, Chang, Xu et~al.}]{vtn}
\bibinfo{author}{Zhao, S.}, \bibinfo{author}{Dong, Y.}, \bibinfo{author}{Chang, E.I.}, \bibinfo{author}{Xu, Y.}, et~al., \bibinfo{year}{2019}b.
\newblock \bibinfo{title}{Recursive cascaded networks for unsupervised medical image registration}, in: \bibinfo{booktitle}{Proceedings of the IEEE/CVF international conference on computer vision}, pp. \bibinfo{pages}{10600--10610}.
\bibitem[{Zhao et~al.(2019c)Zhao, Dong, Chang, Xu et~al.}]{zhao2019recursive}
\bibinfo{author}{Zhao, S.}, \bibinfo{author}{Dong, Y.}, \bibinfo{author}{Chang, E.I.}, \bibinfo{author}{Xu, Y.}, et~al., \bibinfo{year}{2019}c.
\newblock \bibinfo{title}{Recursive cascaded networks for unsupervised medical image registration}, in: \bibinfo{booktitle}{Proceedings of the IEEE/CVF international conference on computer vision}, pp. \bibinfo{pages}{10600--10610}.
\bibitem[{Zhao et~al.(2019d)Zhao, Lau, Luo, Eric, Chang and Xu}]{rcvtn}
\bibinfo{author}{Zhao, S.}, \bibinfo{author}{Lau, T.}, \bibinfo{author}{Luo, J.}, \bibinfo{author}{Eric, I.}, \bibinfo{author}{Chang, C.}, \bibinfo{author}{Xu, Y.}, \bibinfo{year}{2019}d.
\newblock \bibinfo{title}{Unsupervised 3d end-to-end medical image registration with volume tweening network}.
\newblock \bibinfo{journal}{IEEE journal of biomedical and health informatics} \bibinfo{volume}{24}, \bibinfo{pages}{1394--1404}.
\bibitem[{Zhou et~al.(2019)Zhou, Siddiquee, Tajbakhsh and Liang}]{zhou2019unet++}
\bibinfo{author}{Zhou, Z.}, \bibinfo{author}{Siddiquee, M.M.R.}, \bibinfo{author}{Tajbakhsh, N.}, \bibinfo{author}{Liang, J.}, \bibinfo{year}{2019}.
\newblock \bibinfo{title}{Unet++: Redesigning skip connections to exploit multiscale features in image segmentation}.
\newblock \bibinfo{journal}{IEEE Transactions on medical imaging} \bibinfo{volume}{39}, \bibinfo{pages}{1856--1867}.
\bibitem[{Zou et~al.(2022)Zou, Gao, Song and Qin}]{2022review}
\bibinfo{author}{Zou, J.}, \bibinfo{author}{Gao, B.}, \bibinfo{author}{Song, Y.}, \bibinfo{author}{Qin, J.}, \bibinfo{year}{2022}.
\newblock \bibinfo{title}{A review of deep learning-based deformable medical image registration}.
\newblock \bibinfo{journal}{Frontiers in Oncology} \bibinfo{volume}{12}, \bibinfo{pages}{1047215}.

\end{thebibliography}



\end{document}